\documentclass[onecolumn,12pt,showpacs,preprintnumbers,amssymb]{revtex4}

\usepackage{graphics}
\usepackage{graphicx}
\usepackage{epsfig}

\def\anderson{A02}
\def\rbf{{\bf r}}
\def\vbf{{\bf v}}
\def\abf{{\bf a}}

\begin{document}


\title{Independent Confirmation of the Pioneer 10 Anomalous Acceleration}
 
\author{Craig B. Markwardt,\footnote{Electronic address:
        craigm@lheamail.gsfc.nasa.gov}}
\affiliation{Laboratory for High Energy Astrophysics, NASA Goddard Space 
             Flight Center, Code 662, Greenbelt, Maryland, 20771\\
             and
             Department of Astronomy, University of Maryland, College Park, 
             Maryland, 20742-2421}

\date{\today}


\begin{abstract} 
I perform an independent analysis of radio Doppler tracking data from
the Pioneer 10 spacecraft for the time period 1987--1994.  All of the
tracking data were taken from public archive sources, and the analysis
tools were developed independently by myself.  I confirm that an
apparent anomalous acceleration is acting on the Pioneer 10
spacecraft, which is not accounted for by present physical models of
spacecraft navigation.  My best fit value for the acceleration,
including corrections for systematic biases and uncertainties, is
$(8.60\pm 1.34)\times 10^{-8}$ cm s$^{-2}$, directed towards the Sun.
This value compares favorably to previous results.  I examine the
robustness of my result to various perturbations of the analysis
method, and find agreement to within $\pm 5\%$.  The anomalous
acceleration is reasonably constant with time, with a characteristic
variation time scale of $> 70$ yr.  Such a variation timescale is
still too short to rule out on-board thermal radiation effects, based
on this particular Pioneer 10 data set.
\end{abstract}    


\maketitle

 
 

\section{INTRODUCTION\label{sec:intro}}

Measurements of spacecraft motions in the solar system can be used as
tests of gravitation and relativity.  Recently, Anderson et
al. \cite{anderson98,anderson02} have presented the discovery of an
anomalous effect seen in radio tracking data from the Pioneer 10
spacecraft.  When interpreted as a Doppler shift, this anomalous
effect corresponds to a constant acceleration, directed towards the
Sun, of approximately $(8\pm 1)\times 10^{-8}$ cm s$^{-1}$.  Anderson
et al. (hereafter \anderson) found that the anomalous effect could not
be explained by previously known physics or spacecraft properties.

The discovery of the anomaly has stimulated numerous efforts to
explain it.  Some of the explanations involve ``new physics,'' such as
modified gravity or dark matter, while other explanations invoke a
change in the physical properties of the Pioneer spacecraft, such as a
asymmetric radiation profile.  I have considered a third avenue of
exploration, which is to test the analysis procedure for flaws.  In
this paper, I present an independent analysis of the Pioneer 10
trajectory and search for an anomalous acceleration.

\anderson\ studied radio tracking data from four deep space missions:
Pioneer 10, Pioneer 11, Ulysses, and Galileo.  All four of these
missions showed suggestions of an anomalous acceleration of order
$10^{-7}$ cm s$^{-2}$.  However, \anderson\ considered the
determination of the anomalous acceleration of the Pioneer 10
spacecraft to be the most reliable.  Therefore I have focussed
exclusively on the Pioneer 10 data for my analysis.  Once I had
verified the presence of an anomalous acceleration, I tested the
result for robustness in several different ways.

All of the procedures discussed in this paper were developed by myself
\cite{fdf} and written in the Interactive Data Language (IDL)
\cite{rsinc,cmlibrary}.  During the development I had only minimal
contact with the \anderson\ group authors, and as I detail below,
these contacts had a minimal impact.  Thus, I consider this work to be
an independent test of the analysis by \anderson.  I have analyzed a
subset of the Pioneer 10 tracking data that is available from the
public archives, which is, of course, the same data that \anderson\
used in their analysis.  The time coverage of my analysis (years
1987--1994) is most comparable to that of the original discovery
presented in \anderson.  By necessity, many of the analysis procedures
I developed will be at least similar to those of \anderson, but I
attempt to extend the analysis by considering additional models,
including spacecraft spin, maneuvers, and time-variation of the
anomalous acceleration.

The contents of the paper are as follows.  Section \ref{sec:sc-comm}
briefly describes the Pioneer 10 spacecraft and the Deep Space Network
systems.  Section \ref{sec:dataprep} presents the methods I used to
acquire the data and perform initial filtering, while
Sec. \ref{sec:anal} describes the analysis and modeling techniques
that I employed.  The results of the tracking and uncertainty analyses
are presented in Sections \ref{sec:results} and \ref{sec:uncertain}.
This is followed by a short discussion and conclusion in Sections
\ref{sec:discussion} and \ref{sec:conclusion}.

\section{Spacecraft and Communications\label{sec:sc-comm}}

I provide only a cursory description of the Pioneer spacecraft, and
Pioneer and DSN communications systems here.  I refer the reader to
\anderson, and references therein, for a more complete description.

\subsection{Pioneer 10 Spacecraft\label{sec:sc}}

Pioneer 10 was launched on 2 March 1972 and, after an encounter with
Jupiter, has followed a hyperbolic escape trajectory from the solar
system.  On 1 January 1987, the spacecraft was approximately 40
A.U. from the Sun, and receding with a nearly constant velocity of
12.8 km s$^{-1}$.

The main physical features of the Pioneer 10 spacecraft are the
parabolic high gain antenna, with a radius of 137 cm; the instrument
compartment, which faces the direction of travel; and the two
radioisotope thermoelectric generators (RTGs), which are attached to
the compartment by booms of 300 cm length.  The Pioneer 10 spacecraft
is spin stabilized, and spins at a rate of $\simeq 5$ rpm.  The spin
axis is aligned with the high gain antenna axis, which is designed to
point towards the Earth, opposite the direction of travel.

The transponder on board Pioneer 10 functions in the S-band.  The
uplink signal from Earth is received by the high gain antenna at 2.11
GHz, while the downlink signal to Earth is transmitted at a frequency
close to $\nu_o = 2.292$ GHz.  I have used only data where the
spacecraft communications system was operated in ``coherent'' mode.
In this mode, the spacecraft retransmits a downlink signal,
phase-locked to the uplink, with an exact frequency turnaround ratio
of 240/221.

\subsection{DSN Communications\label{sec:comm}}

Communications with Pioneer 10 are accomplished using the Deep Space
Network (DSN).  The DSN maintains antenna complexes at Canberra,
Australia; Goldstone, California; and Madrid, Spain.  Radio tracking
observations are normally obtained in the course of mission operations
for the purposes of spacecraft navigation.  The two basic types of
tracking are ranging, which directly measures the spacecraft distance
via round trip light travel time; and Doppler, which measures
indirectly the range rate, or relative velocity along the line of
sight.  Tracking passes were obtained more or less regularly spaced
throughout the time range 1987--1994, and within individual years,
although there were apparently some campaigns of more intensive
tracking coverage.

During uplinks, a digitally controlled oscillator (DCO) is programmed
to a precise frequency, which then drives the Exciter Assembly, whose
signal is sent to the transmitting antenna.  The uplink frequency is
typically Doppler-compensated so that the frequency received by the
spacecraft is near 2.11 GHz \cite{doppler compensate}.  On the
downlink leg, after being received at Earth, the Pioneer 10 Doppler
signal is down-converted to a 1 MHz intermediate frequency.  The
Metric Data Assembly is used to accumulate a continuous count of
Doppler cycles at this intermediate frequency, at fixed integration
intervals.  The Doppler count so computed represents the integrated
range rate (i.e. the line of sight change in distance between the
antenna and spacecraft).  These quantities are typically differenced
and, after further straightforward manipulations, produce the mean
Doppler frequency over the integration interval
\cite{anderson02,morabito94,moyer00}.

Since the Pioneer spacecraft was 40--69 A.U. from the Earth for the
period of this analysis, the round trip light travel time ranges from
11 to 16 hours.  Thus, while it is possible for the same station to be
used for both the uplink and downlink legs of the transmission to and
from the spacecraft (known as ``two-way'' Doppler), it is more common
for one station to transmit the uplink leg and a separate station to
receive the downlink leg some hours later, which is known as
``three-way'' Doppler.  As far as the analysis is concerned, both of
these kinds of Doppler tracking data are identical and can be handled
the same way.

For spacecraft ranging, a unique repeating ranging code is modulated
onto the 2 GHz carrier wave.  Upon return from the spacecraft, the
received ranging code is correlated with the transmitted one, and a
range time delay can be computed, modulo the period of the ranging
code pattern.  No reliable range data were available for Pioneer 10,
and so I analyzed only the Doppler tracking data.

\section{Data Preparation\label{sec:dataprep}}

I obtained Pioneer 10 Doppler tracking data from the publicly
accessible NSSDC archive.  The JPL Radio Science group has submitted a
substantial portion of the tracking data to the archive in the form of
digital tapes which must be staged manually onto a computer disk.  I
requested data covering the time period 1987--1994.  Data from beyond
April 1994 is not available from NSSDC.  In addition, there are a
number of gaps in the data.  The largest gap is due to an unreadable
archive tape which covered the interval June 1990 to June 1991.  The
data are stored in a standard Archival Tracking Data File (ATDF)
format \cite{trk225}.

An initial level of filtering and processing was applied to the raw
ATDF records.  A large number of records were in ``one-way'' Doppler
mode (i.e., transmissions originating from the spacecraft) and were
simply discarded.  The integration time of the records was variable,
and ranged from 0.1 s (so-called ``high rate'' Doppler), to $\sim 100$
s or more.  The high rate data in particular contained a large number
of samples, and to prevent over-weighting of those segments, I chose
to accumulate the Doppler counts to intervals of at least 60 s in
duration.  I also eliminated discontinuous or noisy data, which occur
preferentially at the beginnings and ends of tracking passes, or
during noisy passes.  A basic sliding 10-sample median filter was
applied, and points more than 100 Hz from the median were discarded.

\section{Analysis Overview\label{sec:anal}}

The data consist of a time series of observed frequencies at
designated DSN antennae.  The dominant variations observed in the data
are the annual signature of the Earth's motion in the solar system and
the diurnal signature of the Earth's rotation.  The diurnal term
contains both the motions of the receiving and the transmitting
antennae involved in the tracking pass.  Finally, of course, there is
the Doppler variation introduced by the Pioneer 10 spacecraft itself,
which is the sought-after signal.

In this section, I will follow the terminology established by
\anderson\ in identifying epoch of transmission and reception.  The
epoch of transmission from the Earth is $t_1$, the epoch of
interaction of the signal with the Pioneer 10 spacecraft is $t_2$, and
the epoch of reception back at the Earth is $t_3$.  All of these times
are referred to the Barycentric Dynamical Timescale (TDB), which is a
coordinate time at the solar system barycenter.  TDB is also the
effective argument of the JPL planetary ephemerides.  The 3-vectors
$\rbf_1$, $\rbf_2$, and $\rbf_3$ represent the positions of the
corresponding antenna at the corresponding epoch, and $\vbf_1$,
$\vbf_2$, and $\vbf_3$ represent the velocities.  The vector
difference, $\rbf_{12}$, is defined as $\rbf_1 - \rbf_2$.  These
vector quantities are measured in the solar system barycenter frame.

The original station times in the ATDF records are referred to
Coordinated Universal Time (UTC).  When computing Earth rotation and
orientation quantities, the Terrestrial Dynamical Time (TDT) timescale
is used.  Conversion between the UTC, TDT and TDB timescales is
straightforward using standard practices \cite{explanatorysupp,fairhead90}.

The expected frequency at the receiver at time $t_3$ can be expressed
as
\begin{equation}
f_{3e} = \left[(240/221)(f_1 d_{12}) - \eta f_{\rm spin}\right] d_{23} + 
   \Delta\nu_{\rm path},
\label{eqn:doppler}
\end{equation}
where $f_1$ is the uplink frequency at time $t_1$ as measured at the
transmitter, $f_{\rm spin}$ is the spacecraft spin frequency at time
$t_2$, and the ratio $240/221$ is the spacecraft transponder
turnaround ratio (note that $\eta = 1 + 240/221$).  The factors
$d_{12}$ and $d_{23}$ embody the Doppler shifts of the moving
spacecraft and earthbound antennae.  The frequency multiplier on the
uplink leg is
\begin{equation}
d_{12} = {\sqrt{1-|\vbf_1|^2/c^2}\over(1 - \hat{\rbf}_{12}\cdot\vbf_1/c^2)}
         {(1 - \hat{\rbf}_{12}\cdot\vbf_2/c^2)\over\sqrt{1-|\vbf_2|^2/c^2}},
\end{equation}
where the first fraction represents the relativistic Doppler shift due
to the Earth motion, and the second due to the spacecraft motion.  The
unit vector $\hat{\rbf}_{12}$ points from the transmitting station to
the spacecraft, i.e., $\hat{\rbf}_{12} = \rbf_{12}/r_{12}$.  The
downlink factor, $d_{23}$, is constructed in the same fashion, by
substituting $1\rightarrow2$ and $2\rightarrow3$.

The final term in equation \ref{eqn:doppler}, $\Delta\nu_{\rm path}$,
represents additional Doppler effects caused by small effective path
length changes, aside from those due to geometric antenna motions.
Generally speaking, this term can be written as $\Delta\nu_{\rm path}
= -2\ dl/dt \ \nu_o/c$, where $dl/dt$ is the time rate of change of
effective photon trajectory path length along the line of sight.  The
factor of 2 comes from the two legs of the round trip path.

In this paper I consider the effective path length due to the
``Shapiro'' delay \cite{shapiro}.  The Shapiro delay reflects the
extra proper distance traveled by a photon, beyond the classical
geometric distance, in the Sun's gravitational potential, as predicted
by general relativity,
\begin{equation}
l_{\rm shap} = (1+\gamma){GM_\odot\over c^2} 
                 \ln\left[{r_1 + r_2 + r_{12}\over r_1 + r_2 - r_{12}}\right],
\label{eqn:shapiro}
\end{equation}
where $\gamma$ is a parameter of the parameterized post-Newtonian
formulation of gravity \cite{Will93,WillNordtvedt72}.  For general
relativity, $\gamma=1$.  On an annual timescale, the impact parameter
of the photon trajectory increases and decreases, with a minimum
distance of about $8\times 10^6$ km.  Conversion to a Doppler shift is
achieved by numerically differentiating equation (\ref{eqn:shapiro}),
which yields an annual signal with amplitude $\pm 150$ mHz.  As
discussed further below, I do not model the effects of the solar
corona.

The known quantities are the receiver quantities $t_3$, $f_3$, and the
station identification for each Doppler sample.  In order to compute
the expected frequency at the same epoch, all of $\rbf_{\{1,2,3\}}$
and $\vbf_{\{1,2,3\}}$ must be determined.  This is especially
important because even the time of transmission, $t_1$, and hence the
transmission frequency, $f_1$, are not known {\it a priori}.  Starting
from the reception epoch, the spacecraft epoch $t_2$ is determined by
solving the light travel time equation \cite{cmlibrary}, $t_3 - t_2 =
r_{23}/c + (l_{\rm shap})_{23}$ via an iterative process, using the
known trajectory and rotation properties of the Earth, and a trial
trajectory of the spacecraft.  In the same way, the transmission epoch
$t_1$ can be determined.  The ATDF data contains a special record of
the transmitter configurations, including the frequency of the
antenna's DCO, from which the transmitted frequency $f_1$ can be
determined. 

The motions of the Earth center are interpolated from the JPL DE405
planetary ephemeris \cite{de405,cmlibrary}, which is referred to the
axes of the International Celestial Reference Frame (ICRF).  The
position and velocity of Earth stations with respect to the geocenter,
referred to an inertial coordinate system such as the ICRF, must take
into account Earth rotation and the changes in Earth orientation
parameters.  The apparent sidereal time, which is the hour angle of
the Earth referred to an inertial system, is taken from Aoki et al.
\cite{aoki82,explanatorysupp}.  Earth precession and nutation describe
the motion of the Earth rotation axis with respect to the celestial
sphere.  These parameters are determined based on the standard IAU
1976 (precession) and IAU 1980 (nutation) theories, and are expressed
as a function of TDT.  The nutation in obliquity and longitude are
corrected using series provided by the International Earth Rotation
Service (IERS \citep{iers:series}), which determines these angles to
high precision via regular Very Long Baseline Interferometry (VLBI)
observations of distant quasars.  The polar motion with respect to the
International Terrestrial Reference Frame (ITRF), and small variations
in the length of day (i.e., UT1$-$UTC) are also taken into account.
Coordinates of earthbound DSN antennae, referred to the ITRF, are also
known to centimeter precision or better based on VLBI, are taken from
an existing DSN publication \cite{folkner:station} (but see
Sec. \ref{sec:results}).

\subsection{Equations of Motion\label{sec:eom}}

The trajectory of the Pioneer 10 spacecraft must be determined by
integrating the equations of motion over the time interval of
interest, given a trial set of initial conditions.  The equations of
motion I used were
\begin{eqnarray}
d\vbf/dt &=& \abf_N + \abf_S + \abf_P\nonumber\\
d\rbf/dt &=& \vbf
\end{eqnarray}
where $\abf_N$ is due to Newtonian gravity, $\abf_S$ is the
acceleration due to solar radiation pressure; and $\abf_P$ is an
anomalous acceleration term (i.e., that which is not accounted for by
known physics).

\anderson\ considers additional terms for the acceleration which allow
for alternate theories of gravity (their equation 3).  I find that
over the span of the data, these terms are always smaller than
$3\times 10^{-12}$ cm s$^{-1}$, and thus I neglect them for the
purposes of Doppler tracking analysis.  Other accelerations which I
disregard: solar wind pressure ($<10^{-13}$ cm s$^{-2}$); collisions
with interplanetary dust ($<10^{-12}$ cm s$^{-2}$, to heliocentric
radii of 60 A.U. \cite{landgraf02,gurnett97}); and the gravitational
attraction of the Kuiper belt ($<3\times 10^{-10}$ cm s$^{-2}$
\cite{anderson02}).

The Newtonian gravitational acceleration was computed as
\begin{equation}
\abf_N = \sum_j{{G M_j (\rbf_j-\rbf)}\over{|\rbf_j-\rbf|^3}}
\end{equation}
where $M_j$ and $\rbf_j$ are the mass and position of solar system
body $j$, referred to the J2000 coordinate frame.  The bodies included
in the sum were the Sun, moon, and planets, the positions of which
were interpolated from the JPL DE405 planetary ephemeris
\citep{de405,cmlibrary}.

In a manner similar to \anderson, I model the acceleration due to
solar radiation pressure as radially directed outward from the Sun
with a magnitude
\begin{equation}
a_S = {\mathcal{K}f_\odot A_P\over c M_P} 
         \left|{{1\ {\rm A.U.}}\over {\rbf-\rbf_\odot}}\right|^2 \cos\theta
\end{equation}
where $\rbf_\odot$ is the barycentric position of the Sun, and the
other constants in the equation are defined in Table \ref{tab:quant}.
For the Pioneer 10 geometric area I have used the area of the high
gain antenna, which has a radius of 137 cm.  The angle of the antenna
to the Sun, $\theta$, is always less than $1.5^\circ$ for Pioneer 10
after 1987, and here I have approximated it as $\theta = 0^\circ$ with
a loss in precision in acceleration of $<4\times 10^{-12}$ cm
s$^{-1}$.

The Pioneer 10 anomalous acceleration, $\abf_P$, is modeled primarily
as a constant acceleration, $\abf_P = a_P \hat{\rbf}$, where here
$\hat{\rbf}$ is a unit vector pointing from the Sun to the spacecraft.
As noted by \anderson\ (and below), the Doppler tracking data for
Pioneer 10 do not permit one to distinguish between a geocentric or
heliocentric acceleration, so this representation is also equivalent
to an acceleration directed along the Earth-spacecraft line.  As I am
using the ``usual'' sign convention for frequencies and velocities
\cite{usual/jpl}, a negative value for $a_P$ will represent an
apparent acceleration towards the Sun.

I also test the constancy of the acceleration by adding a {\it jerk\/}
term,
\begin{equation}
\abf_P(t) = (a_P(0) + j_Pt) \hat{\rbf} \label{eqn:jerk}
\end{equation}
where $j_P$ is the anomalous jerk, which measures the deviation of the
acceleration from a constant.  This expression can be rewritten as
\begin{equation}
\abf_P(t) = a_P(0) (1 + t/T_{j_P}) \label{eqn:reljerk}
\end{equation}
where $T_{j_P} = j_P/a_P(0)$ represents the timescale over which the
anomalous acceleration changes.  However, since the heliocentric
spacecraft velocity is nearly constant with time (heliocentric radial
velocity range of 13.1--12.6 km, with a mean of 12.8 km s$^{-1}$), the
jerk term is also equivalent to a spatial gradient of the anomalous
force, and equation \ref{eqn:reljerk} can also be rewritten as
\begin{equation}
\abf_P(t) = a_P(0) (1 + r/R_{j_P}) \label{eqn:radjerk}
\end{equation}
where r is the heliocentric distance and $R_{j_P} = j_P/(a_P(0) v_r)$
is the physical distance scale for variations in the acceleration.

\subsection{Spacecraft Maneuvers\label{sec:man}}
The Pioneer 10 antenna is designed to point towards the Earth.  As the
spacecraft moves outward through the solar system, regular maneuvers
must be made to adjust the spacecraft attitude to maintain an
Earth-pointing direction.  The spacecraft has two thruster assemblies
mounted on the rim of the high gain antenna, which are aligned with
the antenna and spin axes.  During the maneuvers, the thrusters
execute several small pulses, with each thruster assembly firing in
opposite directions.  The spin axis is gradually precessed until a
spacecraft feedback loop determines that the antenna axis is again
pointed towards the Earth.  According to \anderson, the maneuver
duration is about 15 minutes.

In principle, the impulses from the thrusters are in opposite
directions, and thus should impart no net change in velocity to the
spacecraft.  In practice, the control of the thruster nozzles is
imperfect, and it is possible that a small velocity change will be
imparted during the maneuver.  In most cases, these velocity
increments or decrements are directly visible in the Doppler tracking
data (see Sec. \ref{sec:results}).  I treated these velocity changes
as adjustable parameters.  For the $j$th maneuver, I modeled the
velocity change as $\Delta\vbf_j = \Delta v_j \hat{\rbf}_j$ where
$\Delta v_j$ is a free parameter and $\hat{\rbf}_j$ is a unit vector
which points from the Earth to the spacecraft at the time of the
maneuver.

The precise epochs of the maneuvers are not easily determined from the
ATDF data available from the NSSDC archive.  In principle these data
should always be available, in a ``high Doppler rate'' mode, since
maneuvers can only be performed during tracking passes.  Unfortunately
very little of the high rate data is present in the archive.  Rather
than guess at the maneuver epochs, I requested and obtained from the
Anderson group a file which contained the epochs of the maneuvers as
used in the CHASMP program \cite{slava}.  However, the velocity
increments and directions were determined by my own independent
analysis.

Since the maneuvers are modeled with a single quantity, they determine
the {\it mean} velocity shift per maneuver.  Shorter time scale
effects, like transient leakage from the thruster nozzles, will not be
modeled.  However, \anderson\ found that transient effects were small,
and I will not model them further.  \anderson\ provides a sample case
of a maneuver from December 23, 1993. This data was also present in
the NSSDC archive files, and I was able to verify that the behavior
was very close to that described by \anderson.


\subsection{Spacecraft Spin\label{sec:spin}}

The downlinked tracking signal is affected by the spacecraft spin
(equation \ref{eqn:doppler}).  The nominal spin period is
approximately 4.4 rotations per minute (rpm), however the actual spin
period has varied between between 4.25 and 4.55 rpm over the time span
of the data considered in this paper.  Like the maneuver data, it
could be possible to determine the spacecraft spin from high rate
Doppler data taken during precession maneuvers.  Because this data was
largely unavailable from the NSSDC archive, I also obtained a file
from the Anderson group which contained a detailed Pioneer 10 spin
period history \citep{slava}.  These spin rate measurements come from
a variety of sources \citep{anderson02}, including the star sensor,
the Imaging Photo Polarimeter, and the Doppler signal from precession
maneuvers. During the analysis, I performed linear interpolation
between tabulated points in the file.

While I did not determine the spin history independently, there are
several mitigating factors.  First of all, a spin rate of 4.55 rpm
corresponds to an effective frequency shift of $\sim$75 mHz.  As I
will show, a signal of this magnitude could in principle be detectable
against the Doppler noise, but it is much smaller by a factor of $\sim
40$ than the signal due to the anomalous acceleration by 1994.
Second, the more relevant quantity is how the spin rate {\it change}
affects the Doppler frequency.  For my data, the frequency shift due
to the spin rate change is only $\pm$ 3 mHz, which is essentially
undetectable.  I checked these results by performing an analysis run
where the spin rate was held fixed at its mean value, and also at
zero, and the changes were negligible.

\subsection{Integration of Equations\label{sec:integrator}}

The equations of motion were integrated using an
Adams-Bashford-Moulton predictor corrector algorithm, based on the
DDEABM \citep{depac} routine of the SLATEC library \citep{slatec}
(translated to IDL \cite{cmlibrary}).  This integrator is of variable
order (up to order 13) and adaptive step size.  I adjusted the error
control parameters so that frequency residuals were less than 0.1 mHz.
The initial conditions were the initial spacecraft position and
velocity referred to the solar system barycenter.


\subsection{Additional Filtering\label{sec:filt}}

Several additional data filtering criteria were applied, which relate
to the effects of the Earth's troposphere and the Sun's corona on the
Doppler signal.  The Earth's troposphere is known to introduce an
additional signal propagation delay on the order of tens to hundreds
of nanoseconds.  This effect is strongly dependent on the elevation
angle of the spacecraft to the horizon.  At low elevation angles, the
secant effect multiplies the tropospheric delay by several times.  In
addition, there are terms in the tropospheric delay which depend both
on the season, and atmospheric conditions at the time of the
observation.  These effects are most readily seen in ranging
experiments were signal delay directly corresponds to range error.
For Doppler tracking data, the tropospheric effect enters more subtly,
as the time derivative of the delay, since during a single tracking
pass the spacecraft's apparent position will generally increase or
decrease in elevation.

\begin{figure} 
\includegraphics[height=\columnwidth,angle=+90]{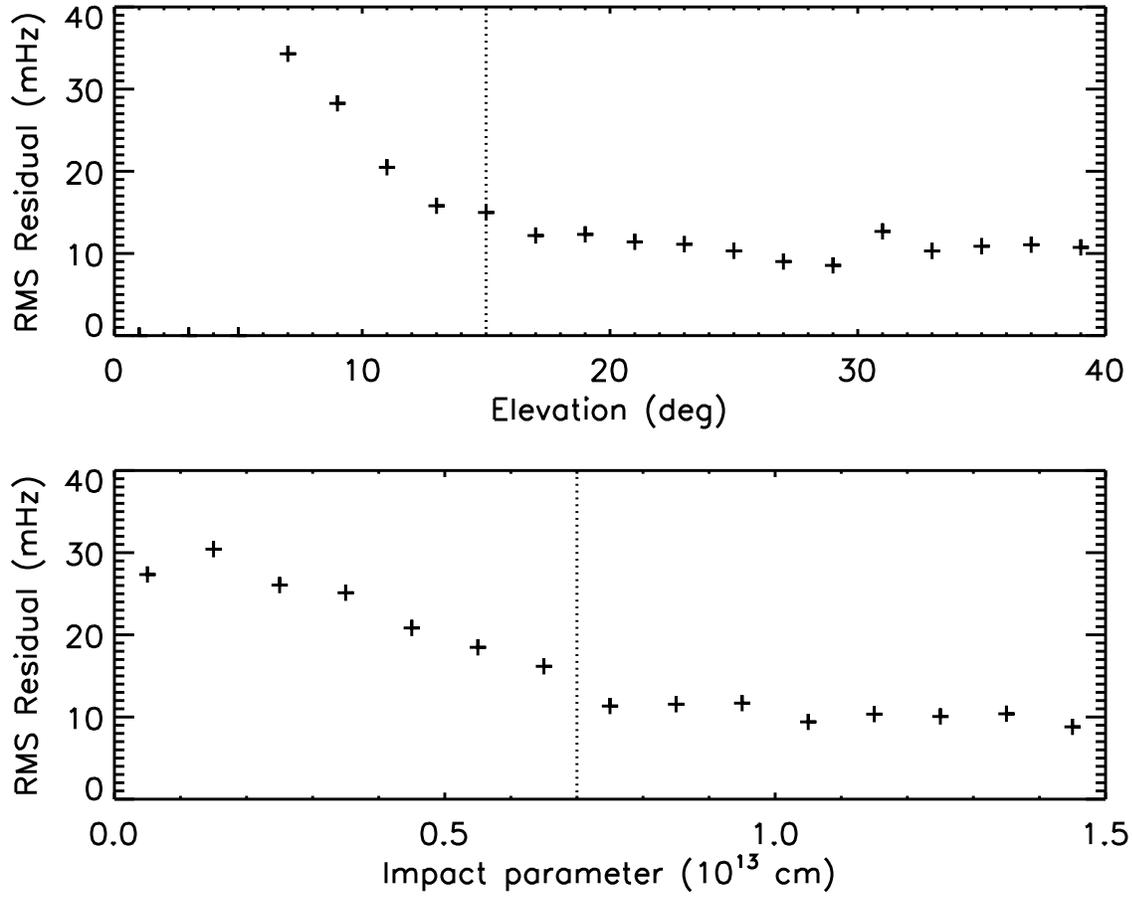}

\caption{Plot of RMS residuals as a function of elevation of
spacecraft above Earth horizon at time of reception (top) and as a
function of point of nearest approach to the Sun of the photon
trajectories (bottom).  Data to the right of the vertical dotted line
were used in the final analysis.
\label{fig:rhel-elv}}
\end{figure}

Using my best-fit model (see below), I divided the residuals into
intervals based on their elevation angle at time of reception.  Figure
\ref{fig:rhel-elv} shows the root mean squared (rms) residuals in each
interval, and demonstrates that below $15^\circ$ elevation there is a
strong increase.  While \anderson\ chose to apply an
elevation-dependent weighting function which included data at low
inclinations, but at a reduced weight, I simply excluded points for
which either the received or the transmitted elevation angle was
smaller than $15^\circ$.

The solar corona also affects the quality of the data.  During solar
conjunctions, the trajectory of photons passes within ten solar radii
of the center of the Sun.  Similar to the troposphere, the bulk solar
corona introduces a variable delay of 0--1.7 $\mu$s.  The derivative
of this variation may be imprinted on the Doppler signal.  However
\anderson\ found that the net effects of the solar corona were small,
and ultimately ignored them.  I constructed a similar coronal model to
that of \anderson\ and also found that the the net effects of the
solar corona on the Doppler signal were small.  However the solar
corona is not a uniform medium.  In addition to the net propagation
delays due to the coronal plasma, there is a general increase in the
Doppler noise. Figure \ref{fig:rhel-elv} also shows a plot of the rms
residuals as a function of the photon trajectory impact parameter.
Clearly the noise is enhanced for impact parameters within $7\times
10^{12}$ cm ($\simeq$ 0.5 A.U.), and so I also elected to exclude any
trajectories which passed within that region.  The excluded segments
are shown in Figure \ref{fig:bestfit}, labeled as ``C.''

Finally, I found that there were several segments of data that were
particularly noisy, and also elected to exclude those from further
analysis.  These segments were from 14 to 29 September 1987, 18
January to 23 February 1992, and 13 March to 29 April 1992, and are
shown in Figure \ref{fig:bestfit}, labeled as ``N.'' I could not find
a direct explanation for why these particular segments were of a lower
quality than the others.

I should note that the exclusion of the segments mentioned above had a
small effect on the result.  When, in a separate analysis, I included
all of the data, the same value for the anomalous acceleration was
reproduced to within 6\%.  However, because of the sensitivity of the
least squares optimization technique to outliers, it is prudent to
exclude highly noisy data which can significantly bias the result.


\subsection{Least Squares Optimization\label{sec:optim}}

The Doppler data were fitted to the model iteratively using a least
squares technique.  The fitting code is based upon MINPACK-1
\cite{minpack,more77}, but translated to IDL \cite{cmlibrary}.  The
free parameters are: (1) the position and velocity of the spacecraft
at the initial epoch; (2) an anomalous acceleration; (3) velocity
increments $\Delta v_j$ due to maneuvers (a total of 18 increments);
and (4) in some cases a jerk term.  The Earth station coordinates and
velocities were also preliminarily considered to be free parameters.
Upon completion of the fit, parameter uncertainties were estimated by
adjusting the Doppler frequency uncertainties so that the $\chi^2$
value was equal to unity, and appropriately rescaling the parameter
uncertainties derived from the covariance matrix of the fit.  In
addition to providing the parameter uncertainties, this method also
provides an estimate of the variance of the Doppler residuals for a
given model.

\begin{figure} 
\includegraphics[height=\columnwidth,angle=+90]{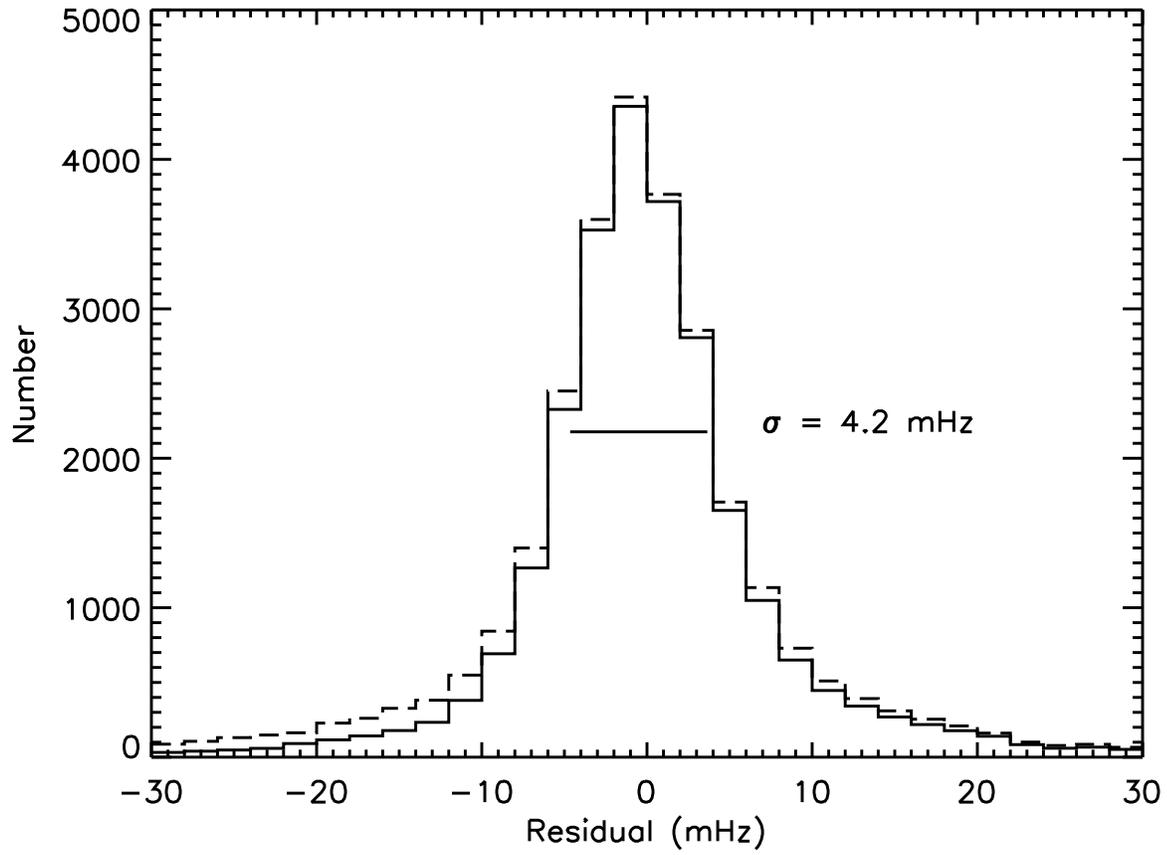}

\caption{Distribution of residuals of the best fit model for only the
filtered data (solid line) and for all of the data (dashed line).  The
curve is characterized by a sharp central peak, well fit by a Gaussian
with the width shown.  The distribution contains significant tails.
\label{fig:resid}}
\end{figure}

\begin{figure}  
\includegraphics[height=\columnwidth,angle=-90]{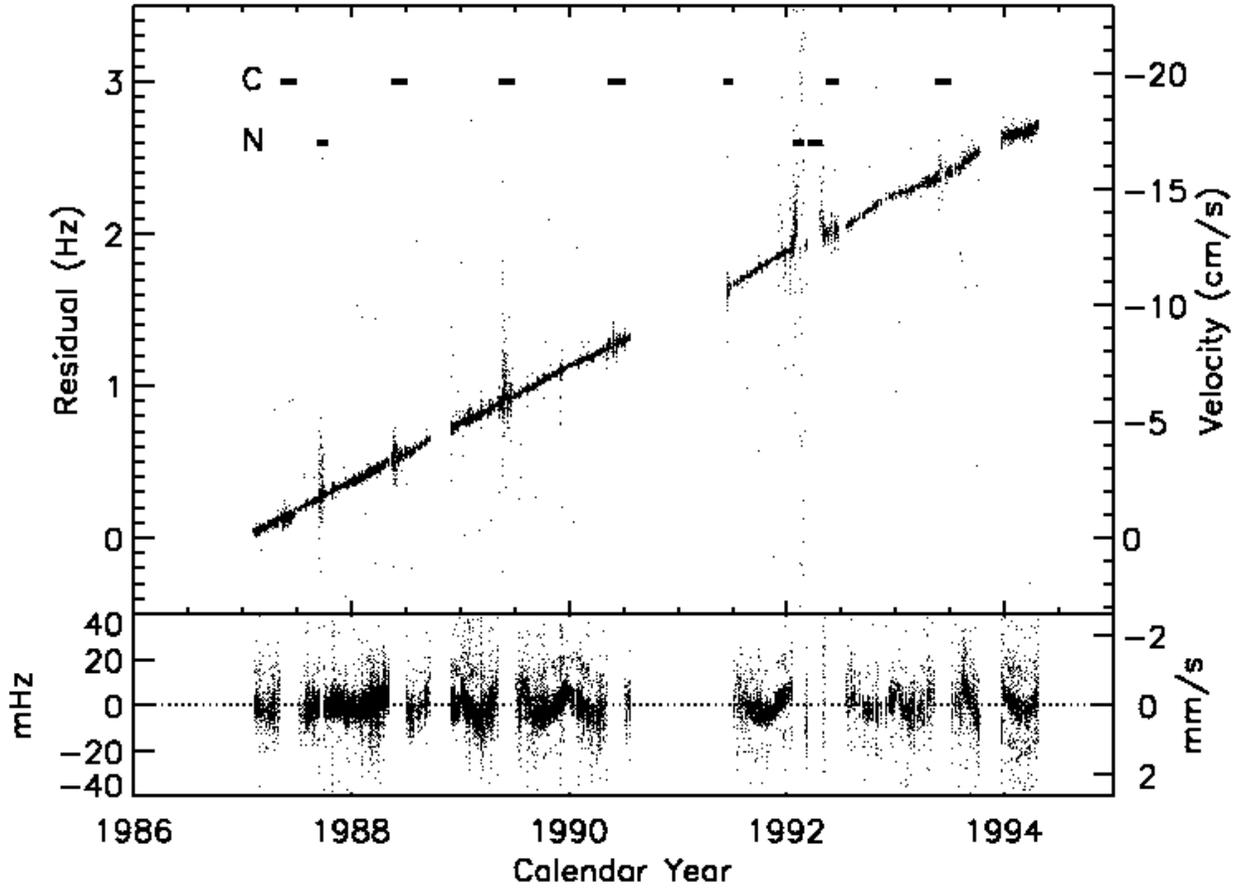}

\caption{Doppler residuals as a function of time of the best fit
model.  The top panel shows the residuals after setting $a_P = 0$, and
demonstrates the linear increase with time.  The top panel shows all
of the data, including segments that were filtered out because of
interference due to the solar corona (designated by a horizontal bar
with ``C'') or due to general noise (designated ``N''; see text).  The
bottom panel shows the filtered residuals, including the best fit
value of the anomalous acceleration.  The equivalent spacecraft
velocity is also shown.  Velocities and frequencies are plotted using
the ``usual'' sign convention \cite{usual/jpl}.
\label{fig:bestfit}}
\end{figure}

Since outliers can still be a problem, I gradually removed the
outliers by applying a threshold filter.  Initially the acceptance
region for residuals was $\pm$10 Hz around zero.  As the fit steadily
improved, I narrowed the acceptance region until I reached a minimum
of $\pm$60 mHz.  The distribution of residuals for the best fit is
shown in Figure \ref{fig:resid}.  The distribution has a clear sharp
peak ($1\sigma$ width of 4.2 mHz), with broad wings that extend at
least to 30 mHz and beyond.  Thus, the measured variance in the
residuals will always be larger than 4.2 mHz, and depend largely on
the size of the acceptance window.  I decided that a $\pm$60 mHz
window was a reasonable compromise between too lax and too aggressive
outlier removal.

A total of 312,116 Doppler records passed the preliminary filtering
process described in Sec. \ref{sec:dataprep}.  I found this number of
data points to be unwieldy to process simultaneously in core memory of
a typical workstation computer, both in terms of memory consumption
and processing time.  I elected to literally decimate the data, taking
only every tenth sample.  This resulted in 31,211 raw records for the
main processing runs.  After application of the corona, tropospheric
and low-noise selection criteria, a total of 23,852 or 76\% of the
records remained.  As a consistency check, I applied the same analysis
to successive independent batches of 31,211 records drawn from the
full pool of Doppler records.  I found that each batch produced
comparable results to the main batch.  The distributions of parameter
values from all ten batches were well matched by the error estimates
taken from the rescaled covariance matrix, and therefore I have
reasonable confidence that the covariance matrix produces appropriate
statistical parameter uncertainties, even in the environment of
outlier points.


\section{Results\label{sec:results}}

In my best fit model I can confirm the signature of a constant
acceleration acting on the Pioneer 10 spacecraft.  Figure
\ref{fig:bestfit} shows the best fit model with and without the
anomalous term \cite{usual/jpl}. Table \ref{tab:anom} shows the best
fitting anomalous acceleration value, $a_P$, for various cases, as
well as the inferred variance in the Doppler residuals.  The best
fitting value of $a_P$ for 1987--1994 is $(7.70\pm 0.02)\times
10^{-8}$ cm s$^{-2}$, where the uncertainty is statistical only.  It
is clear that the Doppler residuals show an increasing trend.  By the
end of the data span in 1994, the frequency of the received Doppler
signal is higher than expected by approximately +2.7 Hz in a single
round trip.

The entry labeled ``\anderson\ Interval I'' refers to the specific
time intervals defined by \anderson.  Interval I spans 3 January 1987
to 17 July 1990 and Interval II spans 17 July 1990 to 12 July 1992.
\anderson\ also considers a third interval which continues up to July
1998.  Because my data set contains a large gap from 1990.5--1991.5,
and no data beyond 1994.3, I consider it inappropriate to quote a
value of $a_P$ in Intervals II and III for direct comparison to
\anderson.  \anderson\ finds anomalous accelerations in Interval I of
$8.02\pm 0.01$ and $8.25\pm 0.02$ for the SIGMA and CHASMP techniques,
respectively ($\times 10^{-8}$ cm s$^{-2}$).  Here I have used the
weighted least squares values with no corona model, as these are the
most comparable to my own.  Generally, there is good agreement between
the work of \anderson\ and myself.


The best fit case was performed without a jerk term.  When a jerk is
included, the fit improves slightly (as judged by the reduction in the
rms residuals), and the anomalous acceleration value increases by
about 5\%.  The small fitted jerk value demonstrates that the
anomalous acceleration is reasonably constant over time, even when
allowing the maneuver parameters to vary.

I have included two fits with simplified spacecraft spin models.  The
first model, ``Mean spin,'' assumes that the spacecraft spin remains
constant at its mean value of 4.40 rpm.  The second model, ``No
spin,'' assumes that the spacecraft has no spin at all.  Both cases
produce results that are essentially indistinguishable from the best
fit case, with similar values of $a_P$ and similar qualities of
residuals.  Thus, while the spin data was not independently determined
by myself, it has little impact on the final result.  The reason that
the zero spin solution does not contain $\sim 75$ mHz residuals is
that these residuals are essentially constant, and can easily be
absorbed into the other free parameters, such as the initial velocity
vector and the maneuver velocity increments.

\begin{figure} 
\includegraphics[height=\columnwidth,angle=-90]{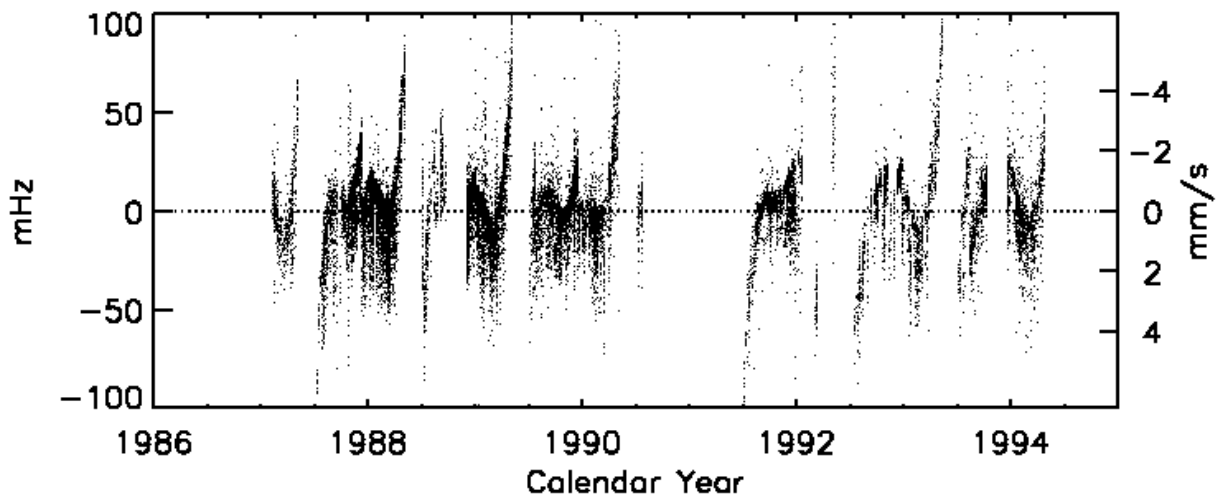}

\caption{Doppler residuals as a function of time of the ``only
maneuvers'' case, showing how the maneuver parameters can absorb some
but not all of the anomaly when $a_P$ is set to zero.  Note the change
in vertical scale from the bottom panel of Figure \ref{fig:bestfit}.
\label{fig:onlymanu}}
\end{figure}

Taking this possibility to its logical extreme, one might surmise that
the entire anomaly could be absorbed into the other free parameters.
The next entry in Table \ref{tab:anom}, ``Only maneuvers,'' fixes
$a_P=0$ while allowing the other parameters to vary.  I found that the
anomaly is indeed absorbed into the maneuver velocity increments, as
might be expected.  However this possibility is not likely for several
reasons.  First, the rms residuals are considerably worse.  If one
were to take the ``best fit'' case as a good fit, i.e. a reduced
$\chi^2$ value of unity, then the ``only maneuvers'' case would have a
reduced $\chi^2$ value of 7.2, which is very unlikely statistically.
The residuals also show systematic trends which actually {\it magnify}
the Doppler discontinuities across maneuver epochs (Figure
\ref{fig:onlymanu}).  Also, one would have to explain how a set of
maneuvers, whose times are irregularly spaced, could produce a steady
increase in the velocity of the craft over 7.5 years to within one
percent.  Finally, the maneuver velocity impulses must be
significantly larger in magnitude than the ``best fit'' case by a
factor of $\simeq 7.5$.  For these reasons I believe that the ``only
maneuvers'' case to be extremely unlikely.

\begin{figure} 
\includegraphics[height=\columnwidth,angle=-90]{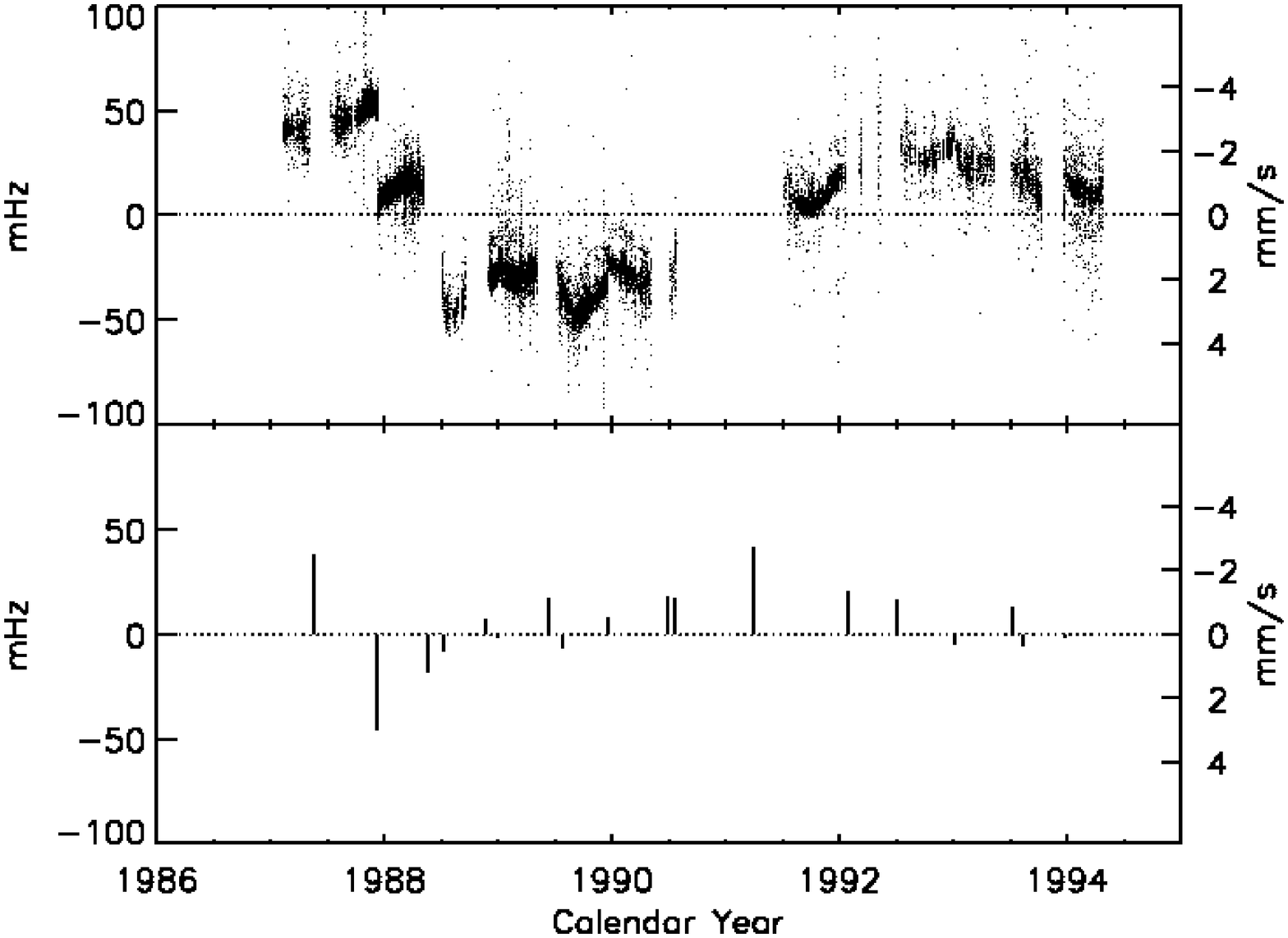}

\caption{Doppler residuals as a function of time for the ``no
maneuvers'' case (top panel), showing that the Doppler shifts of the
maneuvers are visible, but are small in comparison to the overall
anomaly.  The bottom panel shows the fitted velocity increments for
the best fit case.
\label{fig:nomanu}}
\end{figure}

The final entry in Table \ref{tab:anom} is the case where no maneuvers
are modeled, i.e. all of the $\Delta v_j$ are set to zero.  Of course,
Doppler discontinuities are clearly visible in this case (Figure
\ref{fig:nomanu}), but they are (a) small compared to the anomaly, and
(b) both positive and negative sign, compared to the anomaly which is
unidirectional.  This result shows that even without any modeling of
spacecraft maneuvers, the anomaly is significantly detected, and
although considerable effort was put into accurate maneuver modeling,
even a crude model would have sufficed.

In all of the values cited in Table \ref{tab:anom}, the positions and
motions of the Earth stations were fixed to the values determined from
VLBI \cite{folkner:station}.  [The station motions are due primarily
to tectonic drift.]  In another fit (not shown), I allowed the station
coordinates to be free parameters.  I found that the fitted station
coordinates converged to the quoted positions to within a few meters.
Therefore, I left the stations fixed to their fiducial positions.

While \anderson\ divided their data set into three separate intervals,
I do not believe this approach to be appropriate for the abbreviated
data set that I have access to.  Therefore, in my discussion I quote
the ``best fit'' value, which covers the entire 1987--1994 range.

\anderson\ discovered annual and diurnal signatures in their
residuals, which had amplitudes of approximately 10 mHz each.  While
the source was ultimately undetermined, \anderson\ believed the
periodic residuals to be due to previously unrealized errors in the
tabulated solar system ephemerides, and therefore considered it to be
a systematic uncertainty in the analysis.  As can be seen in the
bottom panel of Figure \ref{fig:bestfit}, I also detect modulations of
the $\sim$annual residuals at a similar amplitude.  I also consider
this effect to be a systematic uncertainty.

Finally, I considered the geometric origin of the anomalous
acceleration.  As I have already mentioned, I assumed that the
anomalous acceleration was directed toward the Sun.  In a separate
fit, I adjusted the equations of motion so that the acceleration was
directed toward the instantaneous position of the Earth instead of the
Sun.  This change altered the Doppler residuals systematically by less
than 0.5 mHz, and altered the best fit anomalous acceleration value by
less than 2\%.  Thus, the center of acceleration could be either the
Sun or the Earth and still be consistent with the
data. \cite{heliocentric}.

\section{Anomalous Acceleration and Uncertainties\label{sec:uncertain}}

\anderson\ presented a comprehensive discussion of the systematic
uncertainties associated with the determination of the anomalous
Pioneer acceleration.  I do not intend to repeat such a discussion,
but instead will summarize and adjust it.  \anderson\ divided the
uncertainties into three main categories: those generated external to
the spacecraft, those generated on board the spacecraft, and
computational uncertainties.  

\anderson\ estimated that the uncertainties associated with effects
external to the spacecraft were essentially negligible, with an rms
contribution of $\simeq 0.04\times 10^{-8}$ cm s$^{-2}$.  The largest
estimated systematic uncertainties were associated with effects
generated on board the Pioneer 10 spacecraft.  \anderson\ estimated
the rms contribution of these effects to be $1.27\times 10^{-8}$ cm
s$^{-2}$, which included terms for the reflected heat from the RTGs;
differential emissivity of the RTGs; non-isotropic radiative cooling
of the spacecraft; gas leakage; and other smaller effects.  I adopt
those values here.

The third category, computational uncertainties, were estimated to be
$\simeq 0.35\times 10^{-8}$ cm s$^{-2}$, and included terms for
consistency of modeling ($\sigma_{\text{consist-model}}$) and the
unmodeled annual and diurnal residuals.  \anderson\ was able to rely
on their Interval III (July 1992--July 1998) for the most consistent
determination of anomalous acceleration, but most of that data was not
available to me.  Thus the consistency between different models in my
analysis will by necessity be less.  For the purposes of this work, I
will take $\sigma_{\text{consist-model}}$ to be one half of the range
of anomalous acceleration determinations, or
\begin{equation}
\sigma_{\text{consist-model}} = 0.21\times 10^{-8} {\rm\ cm\ s}^{-2},
\end{equation}
where, to be conservative, I have included the ``extreme'' cases in
Table \ref{tab:anom} (compare to a value of $0.13\times 10^{-8}$ cm
s$^{-2}$ determined by \anderson).  Thus, the total estimated
computational uncertainty is $\simeq 0.38\times 10^{-8}$ cm s$^{-2}$.
The combination of the uncertainties from all three categories,
assuming they are are uncorrelated, is $\sigma_P = 1.34\times 10^{-8}$
cm s$^{-2}$.

\anderson\ also identified experimental ``biases,'' which were other
effects that would tend to systematically increase or decrease the
anomalous acceleration from its experimentally determined value.  For
example, they estimated that the radio transmitter exerts a radiation
force which accelerates the spacecraft at $1.10\times 10^{-8}$ cm
s$^{-2}$, directed away from the Sun.  This acceleration would tend to
{\it increase} the anomalous acceleration.  Their final bias value,
using the sign convention of this paper \cite{usual/jpl}, is $b_P =
-0.90\times 10^{-8}$ cm s$^{-2}$, which I also adopt.

Clearly, the uncertainty in the determination of the anomalous Pioneer
10 acceleration is systematics-dominated and not statistics-dominated.
Determination of the absolute jerk is therefore similarly dominated by
systematic uncertainties.  Formally, I take the upper limit to the
absolute jerk to be
\begin{equation}
|j_P| < \sigma_P/T = 5.7 \times 10^{-17} {\rm\ cm\ s}^{-3},
\end{equation}
where $T$ is the data time span of 7.5 years. This upper limit is a
factor of $\sim$1.5 larger than the value determined in Table
\ref{tab:anom}.  A more interesting quantity is the {\it relative
jerk} (e.g., equation \ref{eqn:reljerk}).  The Doppler tracking data
alone show a reasonably linear correlation with time, and hence
require a small relative jerk.  The effects of a jerk term would be
strongest in the 1994--1998 time range of the \anderson\ data set, but
\anderson\ did not see the effect.  I will therefore still consider
the jerk term shown in Table \ref{tab:anom}, expressed relative to
$a_P$, to be an upper limit.  This leads to
\begin{equation}
|j_P/a_P| = T_{j_P} > 70 {\rm\ yr}
\end{equation}
and
\begin{equation}
R_{j_P} > 170 {\rm\ A.U.},
\end{equation}
which implies that the anomalous acceleration, if it varies, must do
so on broad spatial or temporal scales.  

For my determination of the anomalous acceleration I will assume that
the jerk is zero, and hence use the ``best fit'' case of Table
\ref{tab:anom}.  Following the terminology of \anderson, I label that
experimentally determined quantity to be $a_{P(\text{exper})} =
-7.70\times 10^{-8}$ cm s$^{-2}$.  After adding the bias value and
assigning the systematic uncertainty, I arrive at
\begin{eqnarray}
a_P &=& a_{P(\text{exper})} + b_P \pm \sigma_P \\
    &=& (-8.60\pm 1.34) \times 10^{-8} {\rm\ cm\ s}^{-2}.
\end{eqnarray}

\section{Discussion\label{sec:discussion}}

My best value of the anomalous acceleration agrees quite closely with
the value determined by \anderson\ ($a_{P(\text{\anderson})} =
(-8.74\pm 1.33) \times 10^{-8}$ cm s$^{-2}$), although it should be
pointed out that I have essentially adopted their error analysis
estimates directly. The rms residuals of all of the ``non-extreme''
cases in Table \ref{tab:anom} are of order 8 mHz.  This variance level
is half of the standard error of 15 mHz that \anderson\ assigned to
their Doppler data processing, and thus compares quite favorably with
their result (despite the outliers).

The scope of this paper is to verify the Pioneer 10 anomalous
acceleration by performing an independent analysis.  I will however
discuss briefly some implications for alternate explanations of the
effect.

\anderson\ mentions the Yukawa potential \cite{nieto91/92} as a
candidate form of modified gravity,
\begin{equation}
U(r) = - {{GM}\over{r}}\left({{1 + \alpha e^{-r/\lambda}}\over{1 + \alpha}}\right)
\end{equation}
where $\alpha$ and $\lambda$ are adjustable parameters.  Upon
computing the gradient of this potential and expanding in a power
series of heliocentric radius $r$, one finds
\begin{equation}
a(r) = - {{GM}\over{r^2}} + 
         \left({{\alpha}\over{1+\alpha}}{{GM}\over{2\lambda^2}}\right)
            \left[1 - {2\over 3}{r\over\lambda}\right]
\end{equation}
where the first term is the Newtonian acceleration.  The second term
has a clear analogy to equation \ref{eqn:radjerk}, where $a_P(0)$ is
the term in parentheses, and the length scale $\lambda = 2R_{j_P}/3 >
110$ A.U.  Thus, if the Yukawa acceleration --- or any other
modified-gravity acceleration --- were to deviate from Newtonian plus
a constant, this deviation would occur over spatial scales larger than
the planetary solar system.

The anomalous acceleration has been proposed to be caused by radiation
from the RTGs or electronics in the instrument compartment
\cite{katz,murphy,anderson99,scheffer}.  All electric power on-board
is derived from the RTGs, which in turn derive their power from
radioactive decay of $^{238}$Pu, with a half-life of $\tau = 87.74$
yr.  This radioactivity also produces considerable waste heat of
approximately 2000 W.  As little as 63 W of electromagnetic radiation,
if radiated directionally away from the Sun, could explain the
anomalous effect.  \anderson\ has advanced several
arguments against these classes of explanations for the anomaly.  One
argument is that the anomaly is well enough determined over time that
the radioactive decay of $^{238}$Pu should be detectable, but is not.
If the acceleration were related to heat dissipation, then its
functional form would be
\begin{equation}
a_P = a_P(0) 2^{-t/\tau} \simeq a_P(0) (1 - t \ln 2/\tau).
\end{equation}
This equation is again a direct analog of equation \ref{eqn:reljerk},
however with the constraint that $\tau = T_{j_P} \ln 2 > 50$ yr.  This
limit still accommodates the half-life of $^{238}$Pu, so an explanation
based on radiation from the RTGs cannot necessarily be excluded by my
analysis of the 1987--1994 Pioneer 10 data \cite{electronics}.

If the Doppler errors are considered to be approximately constant over
time, then the error in the jerk should scale as $T^{-2}$, so
additional data over a longer baseline could and should be much more
constraining.  \anderson\ considered the constancy of the anomalous
acceleration by dividing the data into three separate intervals, and
attempting to analyze the intervals independently of one another.
They found a variation of 2.0--5.1\% between their Intervals I and
III.  The ``jerk'' solution I present here would produce a variation
in the acceleration between the midpoints of the two intervals of
about 8\%, a value which is not unreasonably inconsistent with the
results of \anderson.

In order to test the sensitivity to a jerk term, I performed a test
using simulated data.  I used the best fit trajectory {\it with
jerk\/} to construct a synthetic Doppler series, without noise, over
an 11.5 year baseline on a regularly sampled time grid.  I then fitted
that series to a model with {\it no jerk}, but including maneuvers.  I
found that a reasonably good fit could be found.  The rms residuals
were $\sim 1$ mHz, which is much smaller than the typical rms
residuals of the actual best fit models.  The signature of the jerk
was a small parabolic curve in the residuals in each segment between
maneuvers.  Thus, I consider it possible that a jerk term could be
present in the residuals without being readily apparent. An analysis
of the full Doppler data set would be desirable.

\section{Conclusion\label{sec:conclusion}}

I have confirmed by independent analysis that the Pioneer 10 anomalous
acceleration exists in the Doppler tracking data, and is likely not to
be an artifact of the software processing by \anderson.  Direct
comparison to \anderson's SIGMA acceleration value in their Interval I
yields agreement at better than the 1\% level.  The anomaly is robust
to different choices of spacecraft spin model, and also produces a
consistent value even when all maneuvers are removed.  This data does
not constrain whether anomalous acceleration is geocentric or
heliocentric.  By including a jerk term, I have showed that the
acceleration is reasonably constant as a function of time over a 7.5
year time baseline, but not constant enough to rule out thermal
radiation effects due to radioactive decay of Plutonium on board the
spacecraft.

\begin{acknowledgments}

This work uses data provided by the National Space Science Data Center
(NSSDC).  I would in particular like to thank Ralph Post for his
efforts in staging the Pioneer 10 data tapes, and also John Cooper and
Sharlene Rhodes at the NSSDC for their general assistance.  I
appreciate useful conversations with George Dishman.  I thank Slava
Turyshev for providing several data files, and consultation of a
general nature.  I thank Tod Strohmayer and Jean Swank for providing
useful comments on the manuscript.  I acknowledge Aladar Stolmar, who
spurred my initial interest in this subject.

\end{acknowledgments}


\par\vfil\eject


\begin{table}
\caption{Adopted Pioneer 10 and Solar Parameters
\label{tab:quant}}
\begin{ruledtabular}
\begin{tabular}{lc}
Parameter                       & Value\\
\hline
Pioneer 10 Mass, $M_P$ (g)        & $2.51883\times 10^5$ \\
Pioneer 10 Area, $A_P$ (cm$^2$)   & $5.90\times 10^4$ \\
Solar Radiation Constant, $f_\odot$ (erg cm$^{-2}$ s$^{-1}$)
                               & $1.367\times 10^6$ \\
Reflectivity Coefficient, $\mathcal{K}$ 
                               & 1.77 \\
\end{tabular}
\end{ruledtabular}
\end{table}

\begin{table*}
\caption{Pioneer 10 Anomalous Acceleration (Various Procedures)
\label{tab:anom}}
\begin{ruledtabular}
\begin{tabular}{lcc}
Description                               & $a_P$ & RMS residuals\footnotemark[1] \\
                                          & $10^{-8}$ cm s$^{-2}$ & mHz   \\
\hline
Best fit                                  &$-7.70\pm 0.02$       &  7.9  \\
\anderson\ Interval I                     &$-7.98\pm 0.02$       &  7.1  \\
With jerk ($j_P = (+3.7\pm 0.2)\times 10^{-17}$ cm s$^{-3}$)
                                          &$-8.13\pm 0.02$       &  7.8  \\
Mean spin ($f_{\rm spin} = 4.40$ rpm)     &$-7.72\pm 0.02$       &  7.9  \\
No spin ($f_{\rm spin} = 0$)              &$-7.74\pm 0.02$       &  7.9  \\
Only maneuvers ($a_P = 0$)                &0.00\footnotemark[2]  & 21.3  \\
No maneuvers ($\{\Delta v_j\} = 0$)       &$-8.10\pm 0.01$       & 30.2  \\
\end{tabular}
\end{ruledtabular}
\footnotetext[1]{Assumes a window of $\pm 60$ mHz}
\footnotetext[2]{Parameter was fixed}
\end{table*}


\begin{thebibliography}{99}
\bibitem{doppler compensate} The Doppler compensation of the uplink
signal largely accounts for the annual motion of the Earth, for which
only a crude orbit prediction is required.  This compensation does not
disrupt the determination of the anomalous acceleration, since the
relevant quantity is always the {\it difference\/} between the
received and transmitted frequency.
\bibitem{fdf} While spacecraft tracking analysis and orbit
determination is performed by the Flight Dynamics Facility at Goddard
Space Flight Center, I am not affiliated with that group.
\bibitem{usual/jpl} \anderson\ distinguishes between the ``usual'' and
``JPL'' sign conventions for presenting frequencies and velocities.
In this paper, all quantities are presented using the ``usual''
convention.  Higher frequencies are more positive; lower frequencies
are more negative.  A receding spacecraft has a positive velocity; an
approaching one has a negative velocity.
\bibitem{slava} The epochs of precession maneuvers and the spacecraft
spin history were kindly provided by Slava Turyshev of the \anderson\
collaboration.
\bibitem{heliocentric} For completeness I also tested whether the
center of acceleration could be any other planet.  I found that any of
the inner planets (Mecury through Mars) could be the center without
producing significantly different Doppler residuals.  I believe this
is because the inner planets orbit quickly enough that any component
of the acceleration directed away from the sun is averaged to zero
over the Doppler timespan.  If, on the other hand, the center of
acceleration were a fixed point in the solar system, then such
averaging doesn't occur, and the fixed point would therefore need to
be within $\sim 0.5$ A.U. of the Sun to be consistent with the Doppler
data.  I thank Slava Turyshev for suggesting this analysis.
\bibitem{electronics} Similar reasoning can be made regarding
explanations which invoke thermal radiation from the electronic
components, since the electric power output of the RTGs has also been
decreasing on a $\sim$decade timescale (see Figure 16 of \anderson).
\end{thebibliography}
\end{document}